# Tuning the optical forces on- and off-resonance in microspherical photonics

Yangcheng Li[1], Alexey V. Maslov[2], Ana Jofre[1], and Vasily N. Astratov[1,*]

[1] Department of Physics and Optical Science, Center for Optoelectronics and Optical Communication,
University of North Carolina at Charlotte, Charlotte, North Carolina 28223-0001, U.S.A
[2] Department of Radiophysics, University of Nizhny Novgorod, Nizhny Novgorod 603950, Russia
* Tel: 1 (704) 687 8131, Fax: 1 (704) 687 8197
E-mail: astratov@uncc.edu, http://maxwell.uncc.edu/astratov/astratov.htm

**ABSTRACT**
Light pressure effect has been discovered long ago and has been used as an optical method to manipulate micro- and nanoparticles. It is usually considered as a nonresonant effect determined by the transfer of the momentum of light. However, recently we have observed that large polystyrene microspheres of 15-20 μm diameters supporting high quality whispering gallery resonances can be optically propelled in water at an extraordinary high velocity along tapered fibers under resonant conditions. In this work we compare on- and off-resonant optical forces in microspherical photonics by controlling the detuning between the laser emission line and whispering gallery resonances. Our approach involves manipulation with microspheres using conventional optical tweezers and their advanced spectroscopic characterization in fiber-integrated setups. We demonstrate dramatic difference in the optical forces exerted on microspheres in the on-resonant and off-resonant cases. This method can be used to study spectral properties of the resonantly enhanced forces in microspherical photonics.

## 1. INTRODUCTION

Light pressure effect has been discovered long ago and has been used as an optical method to manipulate microparticles. The optical manipulation is usually performed in the near field of a waveguide surface by the means of gradient force and scattering force [1]. Traditionally this effect is considered nonresonant and subtle and determined by the transfer of the momentum of light [2]. It should be noted that although the resonant forces have been brought out by Arthur Ashkin in microdroplets more than 30 years ago [3], the effects observed were relatively weakly pronounced.

However, we have recently observed that large polystyrene microspheres of 15-20 μm diameters can be optically propelled at an extraordinary high velocity along tapered fibers under resonant conditions [4, 5]. Theoretically, we showed that the peak force can approach the level that an absorbing particle would experience based on the total momentum transfer from light to microparticle. Such extraordinary high forces are unusual for transparent dielectric microspheres and they are explained by the efficient transfer of the optical power from the microfiber to the microsphere under resonant conditions. We developed a comprehensive two-dimensional (2D) model of this effect based on a surface electromagnetic wave evanescently coupled to whispering gallery modes (WGMs) in transparent cylinders [5-8]. Experimentally, we observed the strong peaks of the optical forces by studying the instantaneous propulsion velocities of dielectric microspheres along tapered fibers. Giant optical propelling velocities of ~450 μm s$^{-1}$ were recorded for some of the 20 μm polystyrene microspheres in water with guided powers of only 43 mW [5-7]. These experiments were carried out with a flux of microspheres interacting with the tapered fiber. Due to the inevitable ~1% size disorder of microspheres and size-dependent nature of WGM resonances, we realized a random detuning between the laser emission line and WGM resonances in these studies. As a result, the statistical properties of the ensemble of microspheres with random detuning between the laser emission line and WGM resonances were analysed. We concluded that only a small fraction of microspheres interact with the tapered fiber resonantly, and only these microspheres are propelled along the fiber with extraordinary high velocities. Although the experimental results agreed well with that from the theoretical model, it would be desirable to obtain a more direct evidence for the resonant force enhancement based on more controllable detuning between the laser emission line and WGMs in microspheres.

In this work we developed a precisely controlled method to study experimentally the resonant enhancement of optical propulsion force using optical trapping and manipulation of individual microspheres by conventional optical tweezers [9]. We experimentally demonstrated the difference in on-resonant and off-resonant propulsion cases. This method can be utilized to directly prove the role of resonant enhancement of the light pressure effect and further study its properties.

## 2. EXPERIMENTAL PROCEDURES

The essential part of experimental set-up is shown in Fig. 1. A tapered fiber etched from a single mode telecommunication fiber was fixed in a plexiglass platform at the lowest possible position, so that the height of the tapered fiber region above the sealed glass substrate at the bottom of the cuvette was limited at ~100 μm. Since the stable trapping of microspheres could be achieved at the heights limited at ~100 μm in our setup, it was important to position the microfiber below ~100 μm. The tapered region was ~2 mm long and the waist diameter was ~1.5 μm. The platform was filled with distilled water where the tapered fiber was immersed. Optical tweezers function was provided by focusing 1064 nm beam of Nd:YAG laser with ~1W CW power by an Olympus oil-immersed objective with 1.3 NA and 100 magnification. The same objective was used for optical trapping and observation of optical propulsion.

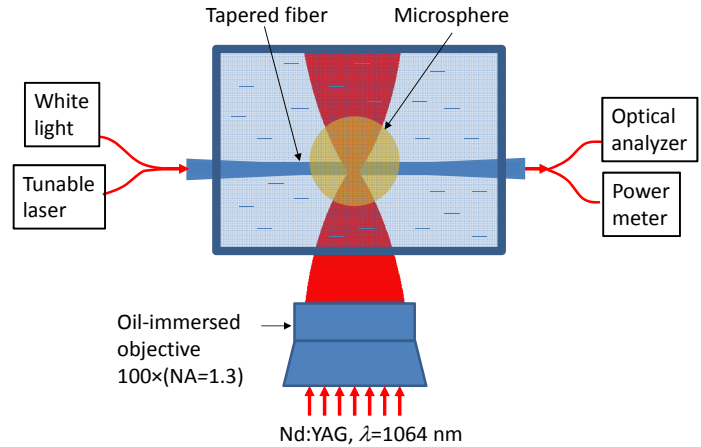

*Figure 1. Experimental setup: microsphere trapped by optical tweezers and brought to the vicinity of the tapered fiber in water immersed integrated platform.*

As illustrated in Fig. 1 individual polystyrene microspheres with 20 μm nominal diameter (Duke Standards[*] 4000 Series Monosized Particles, Thermo Fisher Scientific, Fremont, CA) were trapped by the focused beam and brought to the vicinity of the fiber taper. The input fiber was connected to a broadband white light source (AQ4305, Yokogawa Corp. of America, Newnan, GA) and the output was coupled to an optical spectrum analyzer (AQ6370C-10 Yokogawa Corp. of America, Newnan, GA). Transmission spectra were taken with the sphere hold by the optical tweezers in the near field vicinity of the tapered fiber. It was not possible to independently control the physical contact between the sphere and taper, however due to electrostatic repulsion between them it is likely that a small gap about several tenths nanometers separated the sphere from the taper [6]. The WGM resonances were detected due to dips in the fiber transmission spectra [10]. After taking the transmission spectra the fiber was reconnected to a single mode semiconductor laser (TOPTICA Photonics AG, Gräfelfing, Germany) tunable in 1160–1280 nm range for the optical propelling experiments. The power transmitted through the tapered fiber was varied in 10-30 mW range in different experiments. It was almost two orders of magnitude less compared to the power of conventional optical tweezers used for micromanipulation of spheres.

The laser emission line could be tuned to different wavelengths relative to the WGM resonance dips. When the optical tweezers were turned off, the microsphere would be released and might be propelled along the tapered fiber due to the scattering force exerted by the evanescent field extending outside of the tapered fiber. In this way we can precisely control the resonant conditions under which the tapered fiber interacts with the microsphere and study in detail the resonant enhancement properties of the optical propulsion force.

## 3. EXPERIMENTAL RESULTS AND DISCUSSION

In this work we studied two contrasting cases for comparison: off-resonance, as shown in Fig. 2 (a), when the laser was tuned to the wavelength in the flat transmission region between two adjacent WGM resonances; and on-resonance as shown in Fig. 2 (b), when laser wavelength was tuned precisely in a position of a minimum of WGM resonant dip in the transmission spectrum. After turning off of the optical tweezers, the microsphere would be released and experience the light pressure from the tunable laser. The behavior of microspheres after release was recorded with a CMOS camera. Snapshots were taken from the recorded movies with 100 ms time interval to analyze their motions and calculate applicable propelling velocities.

Two typical scenarios of behavior of spheres corresponding to off-resonant and on-resonant conditions are presented in Fig. 2(c) and Fig. 2(d), respectively. The comparison of these cases demonstrates completely different behavior of spheres in these cases. Under off-resonant condition the propelling effect along the tapered fiber was not observed. As seen in Fig. 2 (c), the sphere slowly drifted away from the fiber most likely due to the fluctuations of the background flux. When the sphere separates from the tapered fiber by more than several microns, it can no longer experience the light pressure that only exists in the near field vicinity of the taper. After separating from the fiber, the spheres slowly descended to the bottom of the cuvette due to the fact that the specific gravity of polystyrene in water is ~1.05. It appears that the gradient force under this condition is not sufficient to trap the sphere around the fiber.

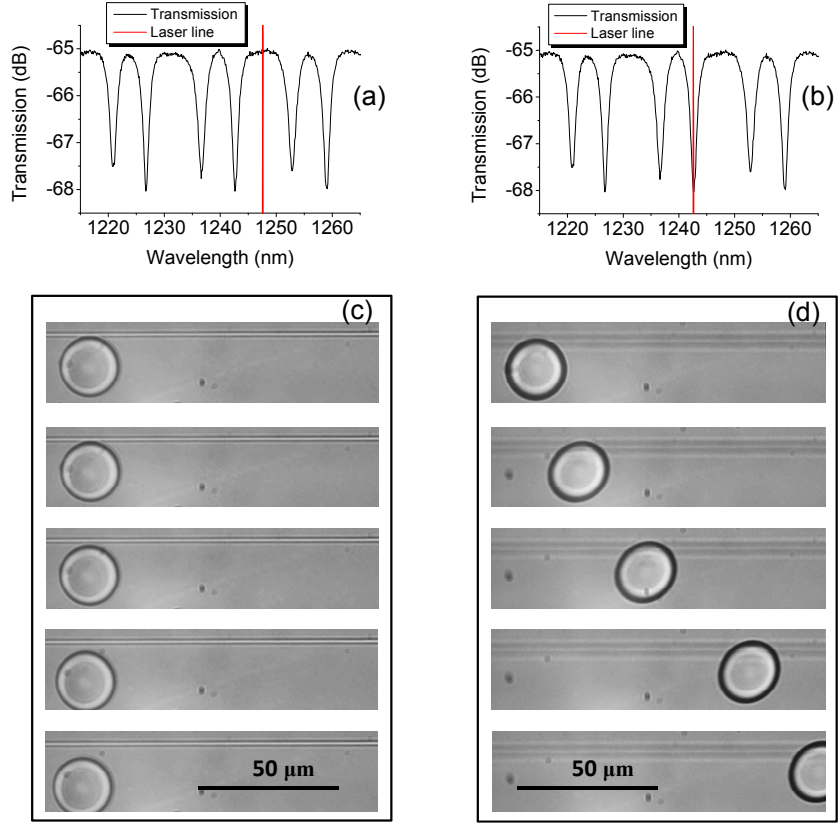

*Figure 2. Transmission spectra of the fiber-microsphere coupler and the laser line for (a) off-resonance and (b) on-resonance cases. Snapshots separated by 100 ms time intervals of the microsphere motions after their release in (c) off-resonance and (d) on-resonance cases. The sphere appears as an ellipse in (d) because of the image distortion introduced by the SIMOS camera for rapidly moving objects.*

In contrast, under on-resonant condition the sphere was found to become radially trapped due to a combination of a long-range attractive interaction and a short-range repulsive interaction. The attractive force is similar to the gradient force in optical tweezers experiments, whereas the repulsive force appears to be due to the repulsion between ionized silanol groups on the bare silica surface, and the negatively charged polystyrene microspheres [6,12]. These results indicate that the radial trapping has a resonant nature determined by the overlap of the laser emission line and the WGM-determined coupling dip in the fiber transmission spectrum. Radially trapped microspheres become a subject for a scattering force directed along the fiber. The particles quickly reach a terminal velocity when the scattering force is equal to the drag force in water. As we showed previously [4-8], the instantaneous velocity can approach extraordinary high values determined by the momentum conservation law for totally light absorbing particles. In this work, we observed instantaneous velocities ~340 $\mu m \ s^{-1}$ with a guided power limited at 26 mW. Normalized by the power, the highest velocity equals ~13 $mm \ s^{-1} \ W^{-1}$ which exceeds our previous observation in Ref. [6] by 30%. Higher propelling velocity is achieved because of the realization of precise resonant conditions in this experiment, compared to the fiber-microsphere interactions with random detuning between the laser and WGMs resonances in our previous studies. With this technique we can make sure that the laser light couples to WGMs in microspheres with the maximal efficiency and thus the propelling force determined by the transferred light momentum is maximal.

## 4. CONCLUSIONS

In this work we performed optical propulsion experiments with controllable detuning between the laser emission line and the WGM-determined resonances. The different behaviors of microspheres under on-resonant and off-resonant conditions were clearly demonstrated. We proved the existence of the resonant enhancement of the radial trapping force and observed the highest instantaneous propelling velocity ~13 mm s$^{-1}$ W$^{-1}$ for 20 μm polystyrene spheres that exceeds all previous results. These effects can be used for sorting microparticles with WGMs peaks resonant with the wavelength of the laser source. Microspheres with resonant WGMs can be used as building blocks of delay lines [13], spectral filters, laser-resonator arrays [14], waveguides [15], focusing devices [16,17], microspectrometers [18], and sensors.

## ACKNOWLEDGEMENTS

The authors gratefully acknowledge support for our work from the US Army Research Office (ARO) under grant W911NF-09-1-0450 and DURIP W911NF-11-1-0406 and W911NF-12-1-0538 (John T. Prater), and from the National Science Foundation (NSF) under grant ECCS-0824067. A.V.M. gratefully acknowledges a partial support from the Ministry of Education and Science of the Russian Federation, agreement no. 14.B37.21.0892.

## REFERENCES

[1] R. F. Marchington, *et al*.: Optical deflection and sorting of microparticles in a near-field optical geometry, *Opt. Express*, vol. 16, pp. 3712-3726, Mar. 2008.
[2] A. Ashkin: Acceleration and Trapping of Particles by Radiation Pressure, *Phys. Rev. Lett.*, vol. 24, pp. 156-159, Jan. 1970.
[3] A. Ashkin and J. M. Dziedzic: Observation of Resonances in the Radiation Pressure on Dielectric Spheres, *Phys. Rev. Lett.*, vol. 38, pp. 1351-1354, June 1977.
[4] Y. Li, *et al*.: Evanescent light coupling and optical propelling of microspheres in water immersed fiber couplers, in *Proc. SPIE*, vol. 8236, Feb. 2012, paper 82361P.
[5] Y. Li, *et al*.: Resonant optical propelling of microspheres: A path to selection of almost identical photonic atoms, in *Proc. ICTON 2012*, Coventry, England, July 2012, paper Tu.A6.2.
[6] Y. Li *et al*.: Giant resonant light forces in microspherical photonics, *Light: Science & Applications*, vol. 2, e64, April 2013.
[7] Y. Li, *et al*.: Giant resonant light forces in microspherical photonics, in *Proc. of CLEO 2013*, San Jose, U.S.A., June 2013, paper CW3F.6.
[8] A. V. Maslov, V. N. Astratov, and M. I. Bakunov: Resonant propulsion of a microparticle by a surface wave, accepted to *Phys. Rev. A* (2013).
[9] A. Ashkin, *et al*.: Observation of a single-beam gradient force optical trap for dielectric particles, *Opt. Lett.*, vol. 11, pp. 288-290, May 1986.
[10] O. Svitelskiy, *et al*.: Fiber coupling to BaTiO$_3$ glass microspheres in an aqueous environment, *Opt. Lett.*, vol. 36, pp. 2862-2864, Aug. 2011.
[11] V.N. Astratov: Fundamentals and Applications of Microsphere Resonator Circuits, in *Photonic Microresonator Research and Applications*, I. Chremmos, O. Schwelb, and N. Uzunoglu, Eds., New York: Springer Series in Optical Sciences, vol. 156, 2010, pp. 423-457, ISBN: 978-1-4419-1743-0.
[12] S. Arnold, *et al*.: Whispering gallery mode carousel – a photonic mechanism for enhanced nanoparticle detection in biosensing, *Opt. Express*, vol. 17, pp. 6230-6238, April 2009.
[13] Y. Hara, *et al*.: Heavy photon states in photonic chains of resonantly coupled cavities with supermonodispersive microspheres, *Phys. Rev. Lett.*, vol. 94, 203905, May 2005.
[14] V.N. Astratov and S.P. Ashili: Percolation of light through whispering gallery modes in 3D lattices of coupled microspheres, *Opt. Express*, vol. 15, pp. 17351-17361, Dec. 2007.
[15] V.N. Astratov, J.P. Franchak, and S.P. Ashili: Optical coupling and transport phenomena in chains of spherical dielectric microresonators with size disorder, *Appl. Phys. Lett.*, vol. 85, pp. 5508-5510, Dec. 2004.
[16] A. Darafsheh, *et al*.: Contact focusing multimodal microprobes for ultraprecise laser tissue surgery, *Opt. Express*, vol. 19, pp. 3440-3448, Feb. 2011.
[17] A. Darafshesh and V.N. Astratov: Periodically focused modes in chains of dielectric spheres, *Appl. Phys. Lett.*, vol. 100, 161121, Feb. 2012.
[18] G. Schweiger, R. Nett, and T. Weigel: Microresonator array for high-resolution spectroscopy. *Opt. Lett.*, vol. 32, pp. 2644-2646, Sept. 2007.